# A Machine-learning Framework for Acoustic Design Assessment in Early Design Stages

Reyhane Abarghooie[1], Zahra Sadat Zomorodian[1], Mohammad Tahsildoost[1], Zohreh Shaghaghian[2]
[1]Shahid Beheshti University, Tehran, Iran
[2]Texas A&M University, College Station, United States

## Abstract

In time-cost scale model studies, predicting acoustic performance by using simulation methods is a commonly used method that is preferred. In this field, building acoustic simulation tools are complicated by several challenges, including the high cost of acoustic tools, the need for acoustic expertise, and the time-consuming process of acoustic simulation. The goal of this project is to introduce a simple model with a short calculation time to estimate the room acoustic condition in the early design stages of the building. This paper presents a working prototype for a new method of machine learning (ML) to approximate a series of typical room acoustic parameters using only geometric data as input characteristics.

A novel dataset consisting of acoustical simulations of a single room with 2916 different configurations are used to train and test the proposed model. In the stimulation process, features that include room dimensions, window size, material absorption coefficient, furniture, and shading type have been analysed by using Pachyderm acoustic software. The mentioned dataset is used as the input of seven machine-learning models based on fully connected Deep Neural Networks (DNN). The average error of ML models is between 1% to 3%, and the average error of the new predicted samples after the validation process is between 2% to 12%.

## Introduction

Acoustic comfort is one of the most important design parameters that affect people's satisfaction and productivity (Claudi, 2019).Therefore predicting the sound conditions in the building is noticeable in both the research and the industrial environments. The quality of the sound environment is linked to numerous physical parameters, which include both the physical properties of sound itself and the physical properties of a room. Sound is characterized by the sound pressure level in a short-term and long-term period and sound frequency (Frontczak & Wargocki, 2011). Physical room characteristics such as geometry, material features, and sound sources influence the acoustic environment.

### Acoustic indicators

A summary of different room acoustic indexes is presented in *Table 1*. Using the following parameters, the acoustic properties of rooms can be identified: reverberation times ($T_{30}$), early decay time (EDT), clarity ($C_{50}$), Definition ($D_{50}$), speech transmission index (STI), percentage articulation loss of consonants (%ALC), etc. (Mikulski & Radosz, 2011) Between the various indicators, T, EDT, $C_{80}$, $D_{50}$, and STI are more frequent among the studies and calculated in most of the acoustic tools.

### Acoustic Simulation Software

*Table 2* summarizes building acoustic software packages that provide some information about room acoustics, noise control, and sound insulations. The information about these tools is collected through manufacturing companies, and among them, Pachyderm is closer to the objectives of this project.

### Calculation Methods

Three techniques are categorized into conventional methods to study sound propagation within the rooms: geometrical models, wave-based models, and statistical models. A comprehensive summary of categories of different acoustic modeling approaches is presented in *Figure 1* (Zhiqiang (John) Zhai, 2016). The parameter values that render the acoustic comfort of a space should be estimated based on the virtual room geometry. Although complete physical models are typically too computationally challenging, simplified mathematical methods such as Sabin or Eyring may be used to create an approximation. The mentioned Formulas can only calculate T, and do not often hold in typical environments such as offices, schools, accommodations, commercials, and healthcare environments.

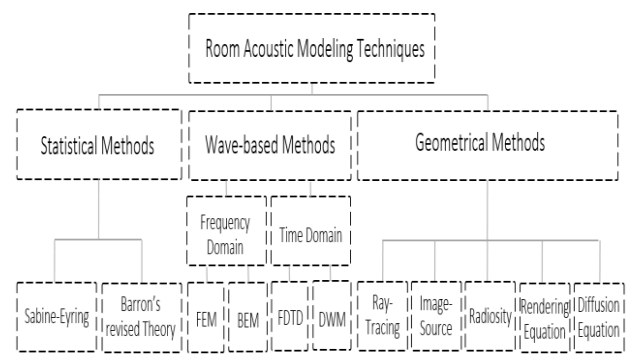

*Figure 1: Categories of different acoustic modeling approaches.* (Zhiqiang (John) Zhai, 2016)

*Table 1: Summary of different room acoustic indexes (Queiroz de Sant'Ana & Trombetta Zannin, 2011; Boulila & Jedidi, 2016; Echenagucia, 2014; Cabrera, 2007; Nowos´wiat & Olechowska, 2016; Navarro Ruiz, 2011; Bowman, 1974; Andersson & Chigot, 2004)*

| Index | Usage | Measurement & simulation | numerical | Description |
|---|---|---|---|---|
| Reverberation Time (T) | Music / Speech | √ | √ | A. It is perceived as the time for the sound to die away. B. More useful among other indexes. |
| Early Decay Time (EDT) | Speech (1 s) / Music (1-2 s) | √ | − | A. Related to the initial and highest level part of the decaying energy. B. Unsuitable for sensitive sound spaces. |
| Clarity ($C_{80}$) | Music (-4 - 2 db) / Speech ( > 3db) | √ | − | A. An early-to-late arriving sound energy ratio. B. mostly used for the musical cases. |
| Definition ($D_{50}$) | Speech (>50%) / Music (<50%) | √ | − | A. Measure of the percent total sound energy arriving within 50 msec. B. Mostly used for the speech cases. |
| Speech Transmission Index (STI) | Speech ( > 0.5) | √ | − | A. STI is the most widely used objective index in usual spaces. |
| Common Intelligibility Scale (CIS) | Speech ( > 0.7) | √ | − | A. It relies on a mathematical relation with STI. |
| Speech Intelligibility Index (SII) | Speech | √ | − | A. Evaluated by speech perception tests given a group of talkers and listeners. |
| Percentage Articulation Loss of Consonants (ALC %) | Speech (< 10%) | √ | − | A. Based on the reception of words by listeners. |
| Center Time (TS) | Speech (60 - 80 ms) / Music (70 - 150 ms) | √ | − | A. Corresponds to the center of gravity of the squared impulse response. |
| Articulation Index (AI) | Speech | √ | − | A. Assess speech intelligibility under a wide range of communication situations. |
| Privacy Index (PI) | Speech | √ | − | A. Based on a mathematical relation with AI. B. It was developed to rate communication rather than privacy. |

*Table 2: Building acoustic software pakcages.*

| software | parameter | Indexes | Advantages | Disadvantages | Online | Free | Type |
|---|---|---|---|---|---|---|---|
| Odeon[1] | room acoustic | SPL, SPL(A), $DL_2$, $T_{30}$, EDT, STI, $T_s$, G, $C_{80}$, $D_{50}$, $LF_{80}$, ST, $LG_{80}$ | import many types of 3d models and Cad / fast / reliable / extendable material library | high-cost | − | − | room |
| SONarchitect[2] | sound insulation | $L_{wA}$ - $R_w$ | easy / fast / visual | − | − | − | building |
| Insul[3] | sound insulation | STC, $R_w$, IIC, $Ln_{Tw}$ | quick / accurate | − | − | − | building |
| Zorba[4] | porous absorption systems properties | α | quick / user friendly | − | − | − | room |
| hvacPRO[5] | mechanical systems noise | − | fast / accurate / high tech / easy to use / adaptable to different needs | − | − | − | building |
| Troldtekt[6] / Jocavi[7] / Dampa[8] | room acoustic | T - SPL | quick / easy / visual | low accuracy / constraints on | √ | √ | room |

---

[1] https://odeon.dk/
[2] https://www.sonarchitect.asia/
[3] http://www.insul.co.nz/
[4] http://www.zorba.co.nz/
[5] https://acousticware.com/
[6] https://www.troldtekt.com/
[7] http://www.jocavi.net/
[8] https://dampa.com/

| | | | | geometry and materials | | | | |
|---|---|---|---|---|---|---|---|---|
| Pachyderm[1] | room acoustic | SPL - $C_t$ - C - $D_{50}$ - EDT - T - STI | open source / simulation of complex geometries / parametric simulation | complicated to use | – | √ | room / building / urban |
| I-Simpa[2] | room acoustic | T - EDT - C - $D_{50}$ - TS - ST | open source | – | – | √ | room / urban |
| Noise 3d[3] | urban noise | SPL | fast / accurate / works on SketchUp | inability to calculate in the frequency range / limited indicators | – | – | urban |
| MithraSIG[4] | urban noise | $L_{aq}$ | quick / accurate / simple modeling / | limited indicators | – | – | urban |
| EXNO[5] | sound insulation of windows | $R_w$ | Revit plug in | limited indicators | – | √ | building |

## Machine Learning

ML is a broad family of techniques, which are often based on statistics, for automatically detecting and utilizing patterns in data. Relative to conventional acoustics and signal processing, ML is data-driven. Given sufficient training data, ML can discover complex relationships between features and desired labels or actions, or between features themselves. ML in acoustics is rapidly developing with compelling results and significant future promise. (Bianco, 2019)

In recent years, limited studies on building acoustic simulation and ML algorithms have been carried out. El-Sallamy, Osman, & Abbas (2019) introduced acoustics software for mechanical engineers and architects to aid in the early design stages or later consultations. Tan (2017) integrated building information modelling (BIM) technology with acoustic simulation to improve indoor acoustic performance. Falcón Pérez (2018) presented an ML-based method to estimate the reverberation time of a virtual room. The main objective of Huang (2017) study was to make a Traffic noise prediction model for high-rise buildings along expressway was constructed with a neural network (NN). Papayiannis, Evers, & Naylor (2019) proposed a method for estimating the probability of the presence of 10 material categories, based on their frequency-dependent absorption characteristics based on ML algorithms. Cao (2020) aims to model the acoustic longevity of asphalt pavement by NN and support vector machines (SVM) algorithms of ML. According to studies, building acoustic tools mostly use traditional methods for calculation, which have some limitations and are not efficient for the first design steps. ML methods have provided significant improvements in data processing and prediction in recent years. But in the field of room acoustics, only one case is mentioned the T index estimation by ML methods. According to *Table 3*, New NN algorithms have produced reliable results in acoustics. To date, there is no room acoustic simulation method that can be reliably used in the early design phases as a decision - making support models for analyzing the room acoustics indicators easily and with high precision and speed.

*Table 3: Acoustic studies by ML algorithms.*

| source | algorithm | number of data | activation function | loss function |
|---|---|---|---|---|
| Falcón Pérez (2018) | NN | 832 | Relu | RMSE - MAE - MAPE |
| Huang (2017) | NN | 327 | Hyperbolic tangent | MSE- CV - MAPE - SE |
| Papayiannis (2019) | NN | 143 | Relu | F1 score |
| Cao (2020) | NN - SVM | 277 | hyperbolic tangent sigmoid | RMSE - MAD - $R^2$ |

## Methods

This research presents a methodology for measuring room acoustic condition in the early design stages of the building by using ML methods. In this way, based on studies, we aim to estimate some of the acoustic indexes, which including $T_{30}$, EDT, STI, $C_{80}$, and $D_{50}$, in typical environments such as offices, schools, accommodations, commercials, and healthcare environments.

### Data

The $T_{30}$, EDT, $C_{80}$, $D_{50}$ values calculate in six typical octave frequency bands (125, 250, 500, 1000, 2000, and 4000 Hz), and the STI index computes in general. The data arise from acoustical simulations of full 3D geometrical modeling of a shoebox room which its furniture has been simplified, and the absorption coefficient of the components are considered.

---

[1] https://www.food4rhino.com/
[2] http://i-simpa.ifsttar.fr/
[3] https://noise3d.com/
[4] https://www.geomod.fr
[5] https://apps.autodesk.com/

## Model Variables

The shoebox room itself is rectangular, mostly empty, and has a window on one side. In the simulation process, variables that include room dimensions, window to wall ratio (WWR), material absorption coefficient, furniture, and shading type have been analyzed. A visualization of eight simulation variables is shown in *Figure 2*. Each arrangement has a particular number, orientation, and position of the items mentioned above. Therefore, there are 2916 different room arrangements available. The data is gathered from each room configuration and from five indexes per frequency, resulting in a total of 72900 data.

| Variables | | | | 2916 |
|---|---|---|---|---|
| Room dimensions | 3x4x3.5 m | 6x7x3.5 m | 8x10x3.5 m | 3 X |
| WWR | 20% | 50% | 80% | 3 X |
| Interior shading | none | Roller blind | curtain | 3 X |
| Furniture | (20% of Floor area) | (40% of Floor area) | | 2 X |
| Wall Absorption coefficient | Gypsum | Wooden | Acoustic Coating | 3 X |
| Floor Absorption coefficient | Ceramic | Parquet | Carpet | 3 X |
| Ceiling Absorption coefficient | Concrete | Gypsum | Acoustic Tile | 3 X |
| Window Absorption coefficient | Single glazed | Double glazed | | 2 |

*Figure 2: Simulation variables. The area of furniture surfaces assumed a percentage of floor area. It was tested some positions before starting the parametric simulation for items of furniture, and then place it in the situation that showed the average of room condition.*

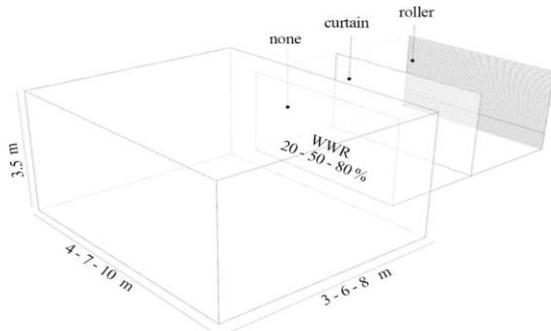

*Figure 3: Shoebox model shape*

## Simulation

Pachyderm plugin has been chosen to use for this project because of its features to calculate the various indicators, the ability of parametric simulation, and its accuracy. Parametric simulation for the shoebox models lasted 45 days due to the time consuming acoustic ray tracing calculations, and the simulation specifications are given in *Table 4*. It is assumed that the background noise is low and following NC30. The sound source is positioned in the corner of the room, and its power is taken based on ordinary speech. For the receiver, it was tested some positions before starting the parametric simulation, then put the receiver in a situation that showed the average of room conditions. The absorption coefficient for all materials in each frequency has been selected following the values provided in the National Building Regulations of Iran, and for the scattering coefficient, the recommended values in Odeon software have been used.

*Table 4: Simulation specifications*

| Calculation method | Ray Tracing / Image source |
|---|---|
| Number of rays | 10000 |
| Cut of time | 1000 |
| Air temperature (C) | 20 |
| Relative humidity (%) | 50 |
| Static air pressure (hPa) | 1000 |

## Deep Neural Network

Deep Learning (DL) algorithms provide an intelligent workflow in which the system can learn from sequential training experiments. (Shaghaghian & Yan, 2020) ML algorithms are powered by data, and the performance of these algorithms on any given task is limited by the quality and quantity of this data. (Falcón Pérez ,2018) The mentioned dataset is used as the input of machine-learning models based on DNN. The data categorized into seven groups. The first six are related to $T_{30}$, EDT, $C_{80}$, and $D_{50}$ indexes in frequency bands, and the last group is relevant to the STI index. The machine learning models are fully-connected DNNs with 5 to 9 hidden layers and 50 to 150 hidden units for each layer. General specifications of DNN models are summarized in *Table 5*, and *Table 6* represents the optimization specifications of each algorithm.

*Table 5: General specifications for all DNN models*

| Number of training data | 2624 |
|---|---|
| Number of testing data | 292 |
| Activation function (hidden layers) | relu |
| Activation function (last layers) | Sigmoid |
| Dropout layer | + |
| Optimizer | Adam |
| Loss function | MSE – MAE – $R^2$ |

*Table 6: Parameter specification of the developed DNN models.*

| Algorithm | Hidden layers | Hidden units | Batch size | Epochs |
|---|---|---|---|---|
| 125 Hz | 7 | 100 | 128 | 100 |
| 250 Hz | 7 | 100 | 128 | 100 |
| 500 Hz | 5 | 150 | 128 | 100 |
| 1000 Hz | 9 | 150 | 128 | 150 |
| 2000 Hz | 5 | 150 | 128 | 150 |
| 4000 Hz | 5 | 100 | 128 | 100 |
| STI | 5 | 50 | 64 | 50 |

**Validation**

The accuracy of each ML prediction model was investigated for each of the outputs. An unseen dataset with 48 model configurations based on Figure 4 specifications was used to assess the predicting models' performance. In this way, the results are calculated once through the new ML method and once through the Pachyderm simulation tool.

Figure 4: Properties of validation variables.

## Results and Discussion

**Sensitivity Analysis**

To increase the interpretability of the "black box" of the DNN models, a new approach called Shapely Additive Explanations (ShAP) (Rathi, 2020) was applied as it could capture the relative importance of parameters in influencing the interaction between input variables. It helps to remove the features, which have a low impact on the model's predictions, and improve the more significant parts. Figure 5 presents the average of sensitivity analysis for the indicators through frequencies. According to the chart, the wall absorption coefficient and the room dimensions are more effective parameters.

**DNN Models Evaluation**

The DNN models were trained and tested using Keras and TensorFlow libraries in Python. The Mean Squared Error (MSE), Mean Absolute Error (MAE), and Cross-Validation (CV) values of training data for the developed DNNs, which can approximate the mentioned indexes, are presented in Table7 Also, *Figure 6* shows the loss function error and test prediction graphs for the 1000 Hz frequency and STI models based on MSE error. Results show that DNN is an effective tool for evaluating the acoustic conditions of the room and allowing the model to achieve an acceptable degree of generalization.

The validation results are also presented in *Table 8*, which compares DNN algorithms with the simulation results for the new data based on MAE error.

ML models provide low and acceptable prediction errors in the training process. However, the same models provided higher error when predicting new results for samples that had not been experienced before. This can indicate a low number of primary data at the time of training. *Figure 7* presents a comparison of DNN errors and validation data.

*Figure 8* shows the web application framework, which consists of three parts. In the first step, the user enters the necessary information. Then, the ML models process in a few seconds, and at the end, the results are presented with the mentioned accuracy.

DNN models make it possible to calculate more variables than those typically used in traditional models. By developing a more advanced DNN model using a sufficiently comprehensive dataset, it will be possible to create more accurate room acoustic index prediction models. Eventually, in future building construction, these prototypes would be useful.

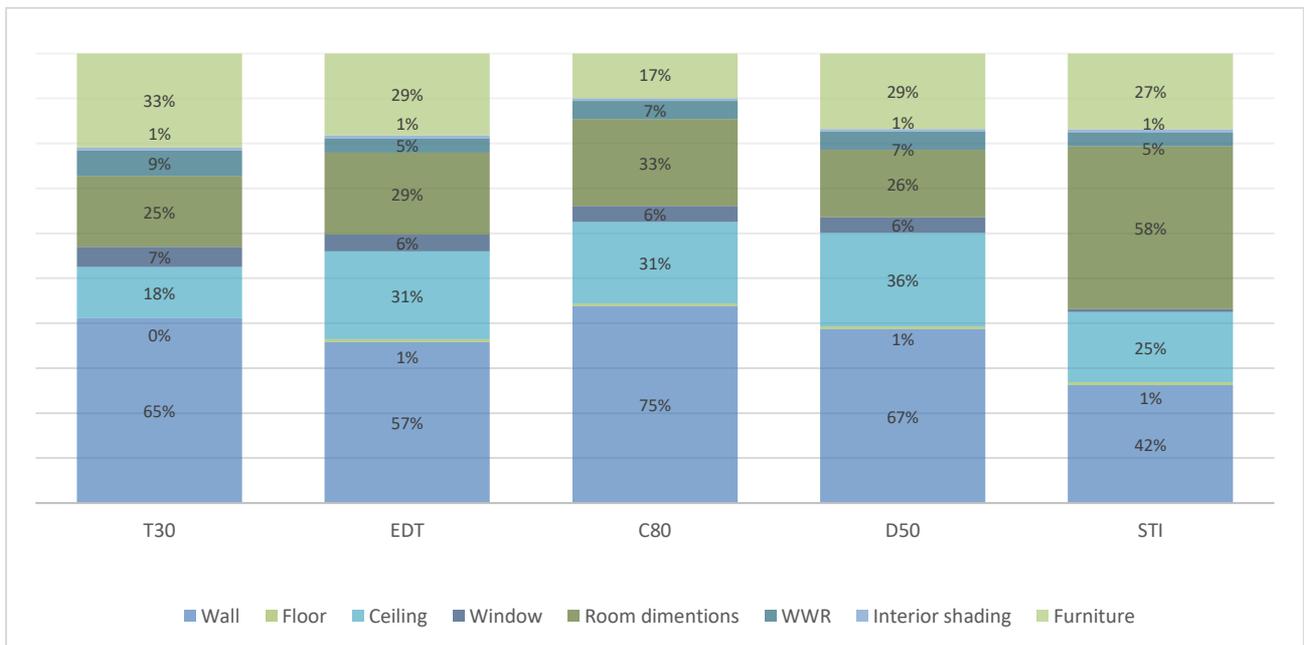

Figure 5: The average of sensitivity analysis for the indicators through frequencies.

Table 7 : The errors of DNN algorithms.

| Algorithms | $T_{30}$ | | | EDT | | | $C_{80}$ | | | $D_{50}$ | | |
|---|---|---|---|---|---|---|---|---|---|---|---|---|
| | MSE | MAE | $R^2$ | MSE | MAE | $R^2$ | MSE | MAE | $R^2$ | MSE | MAE | $R^2$ |
| 125 Hz Frequency | 0.006 | 0.039 | 0.971 | 0.001 | 0.025 | 0.991 | 0.112 | 0.266 | 0.997 | 0.659 | 0.625 | 0.989 |
| 250 Hz Frequency | 0.004 | 0.044 | 0.989 | 0.006 | 0.057 | 0.993 | 0.082 | 0.228 | 0.996 | 0.807 | 0.689 | 0.998 |
| 500 Hz Frequency | 0.006 | 0.036 | 0.988 | 0.001 | 0.028 | 0.995 | 0.073 | 0.212 | 0.998 | 0.541 | 0.567 | 0.994 |
| 1000 Hz Frequency | 0.005 | 0.034 | 0967 | 0.002 | 0.029 | 0.995 | 0.197 | 0.337 | 0.999 | 0.529 | 0.573 | 0.999 |
| 2000 Hz Frequency | 0.020 | 0.077 | 0.935 | 0.001 | 0.021 | 0.995 | 0.47 | 0.459 | 0.998 | 0.784 | 0.625 | 0.998 |
| 4000 Hz Frequency | 0.019 | 0.087 | 0.951 | 0.001 | 0.024 | 0.990 | 0.458 | 0.508 | 0.997 | 0.933 | 0.713 | 0.994 |

| | STI | | |
|---|---|---|---|
| | MSE | MAE | $R^2$ |
| STI | 0.000 | 0.007 | 0.988 |

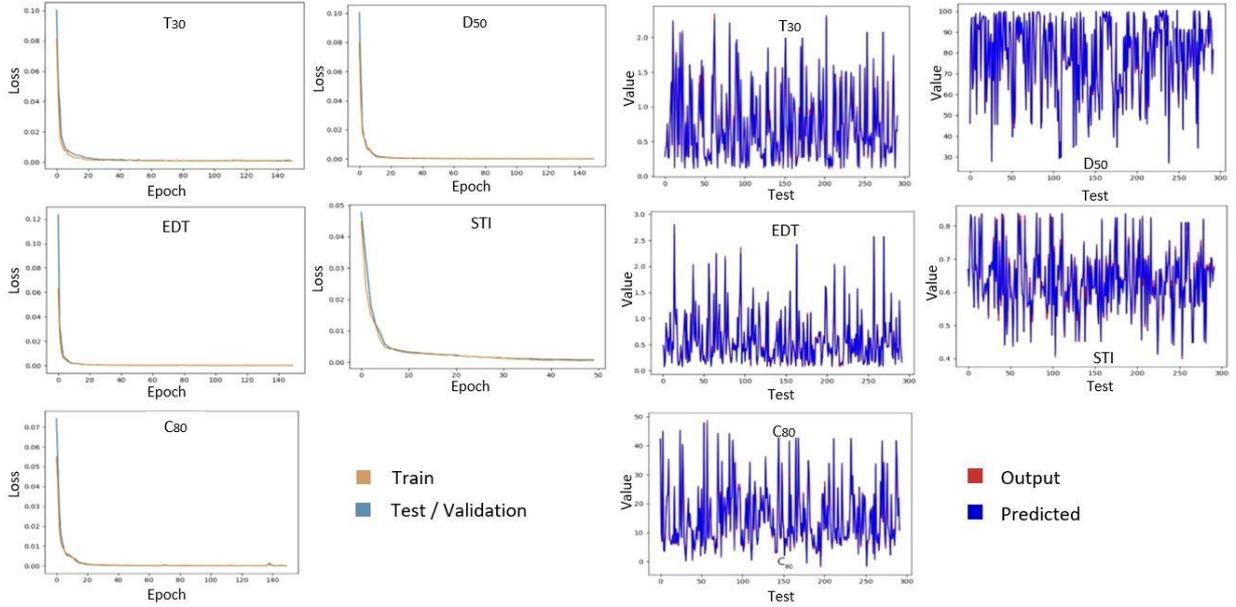

*Figure 6: (left) The loss function error and (right) the test prediction graphs for the 1000 Hz frequency and STI models based on MSE error. The loss function curves show the reduction of errors during the model iterations (the high number of hidden layers in the DNN models has reduced the error of the models quickly), and test prediction graphs show the close prediction of the actual values in the ML models.*

*Table 8: Validation results.*

| Algorithms | $T_{30}$ | | EDT | | $C_{80}$ | | $D_{50}$ | |
|---|---|---|---|---|---|---|---|---|
| | DNN error | New samples | DNN error | New samples | DNN error | New samples | DNN error | New samples |
| 125 Hz Frequency | 0.039 | 0.186 | 0.025 | 0.06 | 0.266 | 1.637 | 0.625 | 3.171 |
| 250 Hz Frequency | 0.044 | 0.148 | 0.057 | 0.053 | 0.228 | 1.226 | 0.689 | 2.850 |
| 500 Hz Frequency | 0.036 | 0.119 | 0.028 | 0.056 | 0.212 | 1.827 | 0.567 | 2.636 |
| 1000 Hz Frequency | 0.034 | 0.338 | 0.029 | 0.088 | 0.337 | 3.772 | 0.573 | 3.973 |
| 2000 Hz Frequency | 0.077 | 0.319 | 0.021 | 0.091 | 0.459 | 6.914 | 0.625 | 4.039 |
| 4000 Hz Frequency | 0.087 | 0.317 | 0.024 | 0.100 | 0.508 | 7.204 | 0.713 | 4.747 |

| | STI | |
|---|---|---|
| | DNN error | New samples |
| STI | 0.007 | 0.025 |

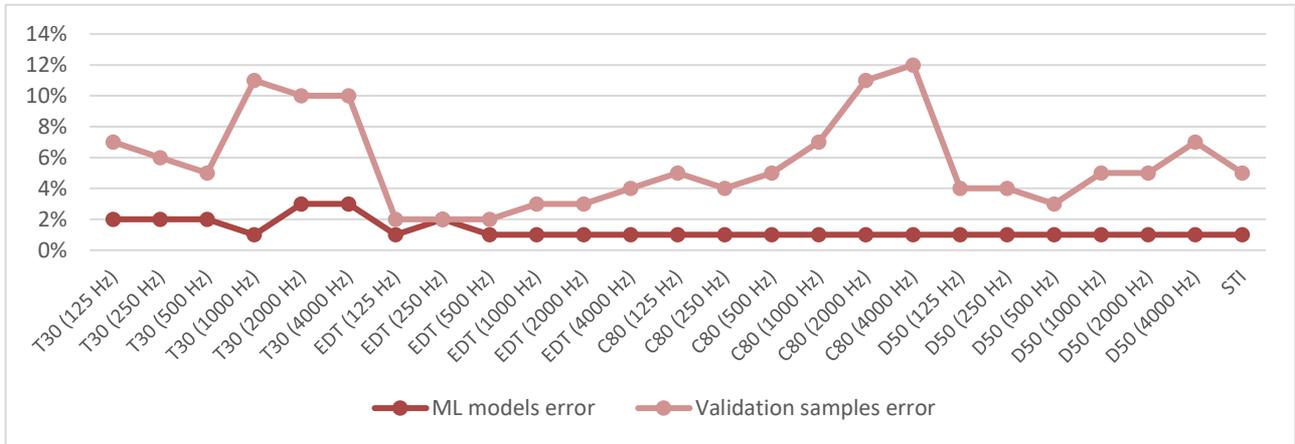

*Figure 7: Percentage of mean error of indicators in machine learning models and validation samples.*

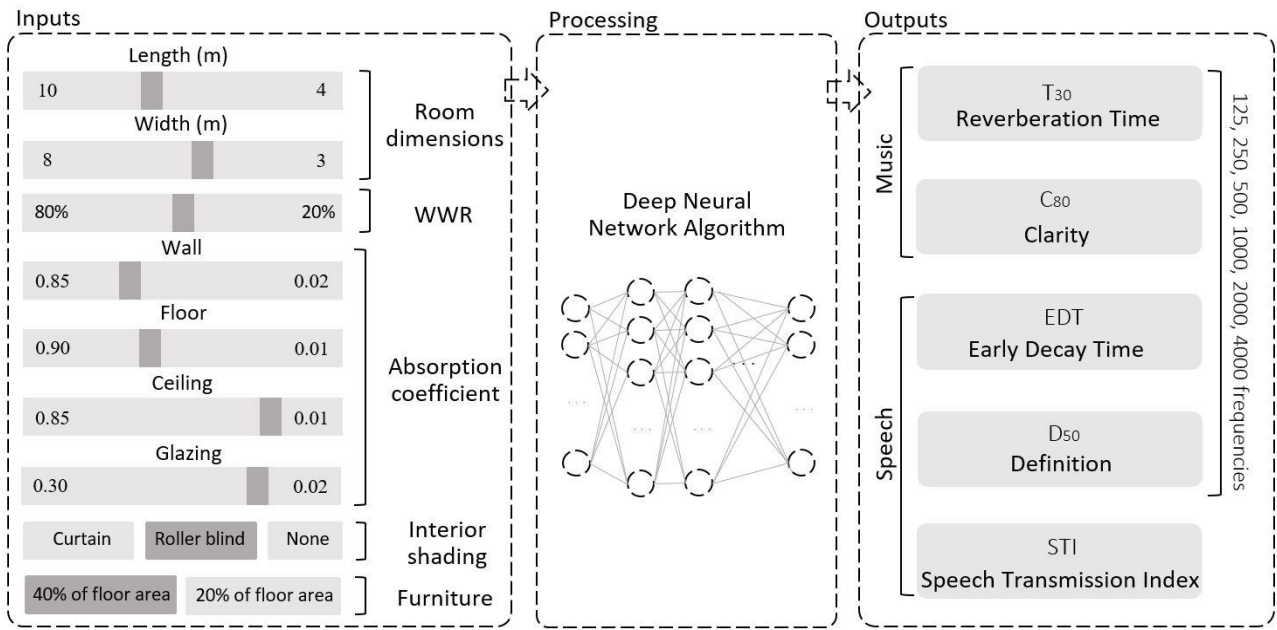

*Figure 8: Application framework.*

While these findings indicate that the current technique can work, there is a need for future work to investigate efficacy under different conditions. The details have to be more diverse, through different room geometries, materials, and furniture details. This research can also be done on sound insulation and noise control to find a simple structure to help designers in the early design stages.

## Conclusion

Acoustic comfort is one of the most important design parameters that affect people's satisfaction and productivity. In this field, building acoustic simulation tools are complicated by several challenges. This work introduces a system focused on machine-learning to estimate the acoustic conditions of the room. The model takes geometric and physical features of a room as input and output the estimated $T_{30}$, EDT, STI, $C_{80}$, and $D_{50}$ indexes values as a frequency function. The calculation was carried out using a DNN-based machine-learning algorithm, and all models were trained and tested in a dataset containing acoustic parametric simulation. According to the results, the accuracy of the models varies in frequencies and for each indicator. The average error of ML models is between 1% to 3%, and the average error of the new predicted samples after the validation process is between 2% to 12%. The findings suggest that the solution proposed can work; moreover, increasing the number of primary datasets can lead to more accurate predicted data.

**Acknowledgement**

The Authors acknowledge Mr. M.J. Hedayati by Road, Housing & Urban Development Research Center for sharing expertise, and valuable guidance.

**References**

Andersson, N.-Å., & Chigot, P. (2004). Is the Privacy Index a good indicator for acoustic comfort in an open plan area? *The 33rd International Congress and Exposition on Noise Control Engineering.* Prague, 22-25 August.

Bianco, M., Gerstoft, P., Traer, J., Ozanich, E., Roch, M., Gannot, S., & Deledalle, C.-A. (2019). Machine learning in acoustics: Theory and applications. *The journal of the acoustical society of America*, 3590–3628.

Boulila, M., & Jedidi, A. (2016). Acoustic study of an auditorium by the determination of reverberation time and speech transmission index. *International Journal of Architectural Engineering & Urban Planning, 26(1)*, 25-32.

Bowman, N. T. (1974). The articulation index and its application to room acoustic design. *Journal of Sound mld Vibration 32(1)*, 109-129.

Cabrera, D. (2007). Acoustic Clarity and Auditory room size perception. *14th International Congress on Sound & Vibration.* Cairns (Australia), 9-12 July 2007.

Cao, R., Leng, Z., Hsu, S.-C., & Hung, W.-T. (2020). Modelling of the pavement acoustic longevity in Hong Kong through machine learning techniques. *Transportation Research Part D 83*.

Claudi, L., Arnesano, M., Chiariotti, P., Battista, G., & Revel, G. M. (2019). A soft-sensing approach for the evaluation of the acoustic comfort due to building envelope protection against external noise. *Measurement*, 675–688.

Echenagucia, T., Sassone, M., Astolfi, A., Shtrepi, L., & Van der Harten, A. (2014). EDT, C80 and G driven auditorium design. *Building Acoustics 21(1)*, 43–54.

El-Sallamy, W., Osman, T., & Abbas, S. (2019). Acoustic consultant – An under-development novel indoors and outdoors noise assessment software. *Ain Shams Engineering Journal 9*, 3305–3313.

Falcón Pérez, R. (2018). Machine-learning-based estimation of room acoustic parameters. *Master thesis. Aalto University.*

Frontczak, M., & Wargocki, P. (2011). Literature survey on how different factors influence human comfort in indoor. *Building and Environment*, 922-937.

Huang, B., Pan, Z., Liu, Z., Hou, G., & Yang, H. (2017). Acoustic amenity analysis for high-rise building along urban expressway: Modeling traffic noise vertical propagation using neural networks. *Transportation Research Part D 53*, 63–77.

Mikulski, W., & Radosz, J. (2011). Acoustics of classrooms in primary schools – Results of the Reverberation Time and the Speech Transmission Index assessments in selected buildings. *Archives of Acoustics*, 777–793.

Navarro Ruiz, J. M. (2011). Discrete-time modelling of diffusion processes for room acoustics simulation and analysis. *PhD thesis. Politécnica de Valencia University.*

Nowos´wiat, A., & Olechowska, M. (2016). Fast estimation of speech transmission index using the reverberation time. *Applied Acoustics 102*, 55–61.

Papayiannis, C., Evers, C., & Naylor, P. (2019). Detecting sound-absorbing in a room from a single impulse response using a CRNN.

Queiroz de Sant'Ana, D., & Trombetta Zannin, P. H. (2011). Acoustic evaluation of a contemporary church based on in situ measurements of reverberation time, definition, and computer-predicted speech transmission index. *Building and Environment (46)*, 511-517.

Rathi, P. (2020). *A novel approach to feature importance — Shapley Additive Explanations*. Retrieved from towards data science: https://towardsdatascience.com/a-novel-approach-to-feature-importance-shapley-additive-explanations-d18af30fc21b [Accessed 19 Dec. 2020]

Shaghaghian, Z., & Yan, W. (2020). Application of Deep Learning in generating desired design options: experiments using synthetic training dataset. *2020 Building Performance Analysis Conference and SimBuild co-organized by ASHRAE and IBPSA-USA.*

Tan, Y., Fang, Y., Zhou, T., Wang, Q., & Cheng, J. (2017). Improve indoor acoustics performance by using building information modeling. *34th International Symposium on Automation and Robotics in Construction.*

Zhiqiang (John) Zhai, H. W. (2016). Advances in building simulation and computational techniques: A review between 1987 and 2014. *Energy and Buildings*.